\documentclass[aps,prd,nofootinbib,twocolumn,floatfix,
showpacs,
letterpaper]{revtex4}
\usepackage{graphicx}\usepackage{natbib}\usepackage{times}\usepackage{amsmath}\usepackage[T1]{fontenc}\usepackage{textcomp}\usepackage{mathptmx}

\begin{document}

\title{Black Hole solutions in a cosmological spacetime background}

\author{Metin Ar{\i}k}
\email{metin.arik@boun.edu.tr}
\author{Yorgo \c{S}eniko\u{g}lu}
\email{yorgo.senikoglu@boun.edu.tr}
\affiliation{Department of Physics, Bo\u{g}azi\c{c}i University, Bebek, Istanbul, Turkey}
\date{January 18, 2014}

\begin{abstract}

We propose and analyze a new metric that has two conformal factors a(t) and b(t) that combine the expansion of the universe and its effects on the spatial and temporal part of the Schwarzschild metric in isotropic coordinates. We present the solutions, their descriptions and we comment on their shortcomings. In the spatially flat case of an expanding universe, we derive from the proposed metric the special solutions of the field equations for the dust approximation and the McVittie metric. We show that the presence of a black hole does not modify the a(t)\hspace{1mm}$\alpha$ $t^{2/3}$ law for dust and $H=constant$ for dark energy.
\end{abstract}

\maketitle

\section{Introduction}
In Einstein's theory of general relativity, two exact solutions were well known and studied throughout the years. One is the Schwarzschild solution that describes the gravitational field outside a spherical non-rotating mass, without charge, and the cosmological constant set to zero. This was practical to model spacetime outside a star, a planet or a black hole.
In Schwarzschild coordinates (on a spherically symmetric spacetime) the line element for the Schwarzschild metric has the form
\begin{multline}
ds^2=(1-\frac{2Gm}{r})dt^2-(1-\frac{2Gm}{r})^{-1} dr^2-r^2 d{\Omega}^2\\where \hspace{2mm}d{\Omega}^2 = d{\theta}^2+\sin^2\theta d{\phi}^2.
\end{multline}

Another formulation commonly used of this spacetime is in the isotropic coordinates.
\noindent
The metric takes the form

\vspace{1mm}
\begin{equation}
ds^2=\frac{(1-\frac{Gm}{2r})^2}{(1+\frac{Gm}{2r})^2}dt^2-(1+\frac{Gm}{2r})^4 (dr^2+r^2 d{\Omega}^2).
\end{equation}
The issue of the Schwarzschild metric is that it ignores the fact the universe is expanding in the background in which the mass is present.
\vspace{2mm}

Another exact solution of the Einstein Field Equations is the Friedmann Robertson Walker Lema\^{i}tre (FRWL) metric\cite{A},\cite{B}, with the assumption that space is homogeneous and isotropic, and the spatial part of the metric may depend on time.

\noindent
For a spatially flat universe, the FRWL metric is

\begin{equation}
ds^2=dt^2-a(t)^2(dr^2+r^2d{\Omega}^2).
\end{equation}
\noindent

\vspace{4cm}
The non-vanishing components of the Einstein tensor in the orthonormal basis are

\begin{equation}
{G}_{00}=3\frac{\dot{a}^2}{a^2}\hspace{1mm},
\end{equation}
\begin{equation}
{G}_{11}={G}_{22}={G}_{33}=-2\frac{\ddot{a}}{a} - \frac{\dot{a}^2}{a^2}\hspace{1mm}.
\end{equation}

Concerning the impacts in an expanding universe,
authors V.Faraoni\cite{C} and A.Jacques\cite{D} worked on  the cosmological effects of expansion on local systems. They explored the local attraction in a gravitationally bound system and  analysed the solution  of general relativity  representing a black hole embedded in a special cosmological background. H.Arakida\cite{E} presented the dominant effects due to cosmological expansion, and the use of a time dependent spacetime model.
 
With a time dependent model McVittie\cite{F},\cite{G}\hspace{1mm}incorporated two solutions to depict a spherically symmetric metric that describes a point mass embedded in an expanding spatially-flat universe.

The McVittie metric is

\begin{equation}
ds^2=\frac{(1-\frac{Gm}{2ra(t)})^2}{(1+\frac{Gm}{2ra(t)})^2}dt^2-a(t)^2(1+\frac{Gm}{2ra(t)})^4(dr^2+r^2 d{\Omega}^2).
\end{equation}

The non-vanishing components of the Einstein tensor in the orthonormal basis are
\begin{equation}
{G}_{00}=3\hspace{1mm}\frac{\dot{a}^2}{a^2}\hspace{1mm},
\end{equation}
\begin{equation}
{G}_{11}={G}_{22}={G}_{33}=-\frac{1}{(1-\frac{Gm}{2ra(t)})} \bigg(2\frac{\ddot{a}}{a} + \frac{\dot{a}^2}{a^2} + \Big(\frac{Gm}{2ra(t)}\big(2\frac{\ddot{a}}{a} -5 \frac{\dot{a}^2}{a^2}\big)\Big)\bigg).
\end{equation}

N.Kaloper, M.Kleban and D.Martin\cite{H} studied thoroughly the McVittie solution that contains a regular black hole in an expanding universe, worked on specific solutions that asypmtote  Friedmann Robertson Walker Lema\^{i}tre (FRWL) universes dominated by a positive cosmological constant and A.M. da Silva, M.Fontanini and D.C.Guariento\cite{I}, the main characteristics of the McVittie solution for different choices of the scale factor.
R.Nandra ,A.N.Lasenby and M.P.Hobson\cite{J} present a tetrad based procedure to solve Einstein's field equations to derive metrics describing a point mass in an expanding universe and B.C.Nolan\cite{K} proves the existence of solutions representing more general spherical objects embedded in a Robertson Walker universe.
F.R.Klinkhamer\cite{N} gives a regularization of the Schwarzschild solution over a simply-connected manifold, which has a curvature singularity at the center.It is shown that spherically symmetric collapse can result in this nonsingular black-hole solution,if the topology near it, is affected by quantum gravity effects.
G.Cristofano, G.Maiella and C.Stornaiolo\cite{O} quantize the entropy of extremal black holes, and extend it to Schwarzschild black holes to obtain interesting cosmological consequences concerning their energy levels spacing. 

This leads us to a very na\"{\i}ve question; why should the conformal factor (scale factor) a(t) influence the same way the temporal part and spatial part of the metric. Can a more general expression of the metric be presented?  The motivation that drove us also to this, was the question whether or not the Schwarzschild Radius was increasing in an expanding universe. This particular question was raised concerning the event horizon of an evolving universe in the paper by A.Davidson and S.Rubin \cite{L}. The authors have proven that any static metric with a Killing horizon in the presence of a perfect fluid, is necessarily a Schwarzschild solution with a vanishing proper energy density and a vanishing proper pressure.
Solutions describing black holes embedded in gradually increasing perfect fluid and the phenomenon for $p=\omega\rho$ were demonstrated. Leading to the proof that the only perfect fluid $p=\omega\rho$ static, spherically symmetric black hole solution is the Schwarzschild solution with vanishing p and $\rho$.
The authors clearly respond to the question "Can an evolving universe host a static event horizon?" by proving that locally the answer is yes with Hawking temperature included.

Thus we wanted to see if a conformal factor can influence with a time dependency, the radius of this black hole and therefore the dynamics that it creates with time. In the following section we will propose more generalized solutions of the Field Equations that lead to interesting findings.

\section{General Formulation}

Let us write the metric with two conformal factors a(t), the scale factor and b(t), a conformal factor.

\begin{equation}
ds^2=\frac{(1-\frac{Gmb(t)}{2r})^2}{(1+\frac{Gmb(t)}{2r})^2}dt^2-a(t)^2(1+\frac{Gmb(t)}{2r})^4 (dr^2+r^2 d{\Omega}^2).
\end{equation}
\noindent
The calculations lead us to the following non-vanishing components of the Einstein tensor in the orthonormal basis

\begin{equation}
{G}_{00}=3\hspace{1mm}\frac{\dot{a}^2}{a^2}\hspace{1mm}\frac{1}{(1-\frac{Gmb(t)}{2r})^2}\hspace{1mm}\big(1+\frac{Gm}{2r}(b+2\dot{b}\hspace{1mm}\frac{a}{\dot{a}})\big)^2\hspace{1mm},
\end{equation}
\begin{equation}
{G}_{01}=\frac{2Gm}{r^2(1-\frac{Gmb(t)}{2r})^2(1+\frac{Gmb(t)}{2r})^2}\frac{(b\dot{a}+\dot{b}a)}{a^2}\hspace{1mm},
\end{equation}
\begin{multline}
{G}_{11}={G}_{22}={G}_{33}=-\frac{1}{(1-\frac{Gmb(t)}{2r})^3}\big(2\frac{\ddot{a}}{a}+\frac{\dot{a}^2}{a^2}\\+\frac{1}{a^2}(\frac{Gm}{2r})(b\dot{a}^2+16a\dot{a}\dot{b}+2a\ddot{a}b+4a^2\ddot{b})\\+\frac{1}{a^2}(\frac{Gm}{2r})^2(-b^2\dot{a}^2+4ab\dot{a}\dot{b}-2ab^2\ddot{a}+16a^2\dot{b}^2)\\+\frac{1}{a^2}(\frac{Gm}{2r})^3(-b^3\dot{a}^2-12ab^2\dot{a}\dot{b}-2ab^3\ddot{a}-8a^2b\dot{b}^2-4a^2b^2\ddot{b})\big).
\end{multline}
All the equations (10) to (12) are satisfied.
\section{Discussions of Solutions}
\subsection{Dust approximation}

Matter dominated Universe is modeled by dust approximation. The matter is approximated as stationary dust particles which produce no pressure.

\vspace{2mm}
\noindent
For {\it{p}} = 0, 
\begin{equation*}
{G}_{11}={G}_{22}={G}_{33} =0. 
\end{equation*}
\noindent
Equating (12) to zero, we obtain a system of differential equations.
\begin{equation*}
2\frac{\ddot{a}}{a}+\frac{\dot{a}^2}{a^2}=0\hspace{1mm},
\end{equation*}
\begin{equation*}
(b\dot{a}^2+16a\dot{a}\dot{b}+2a\ddot{a}b+4a^2\ddot{b})=0\hspace{1mm},
\end{equation*}
\begin{equation*}
(-b^2\dot{a}^2+4ab\dot{a}\dot{b}-2ab^2\ddot{a}+16a^2\dot{b}^2)=0\hspace{1mm},
\end{equation*}
\begin{equation*}
(-b^3\dot{a}^2-12ab^2\dot{a}\dot{b}-2ab^3\ddot{a}-8a^2b\dot{b}^2-4a^2b^2\ddot{b})=0.
\end{equation*}
The unique solution that satisfies this system of equations is
\begin{equation*}
a(t) = ct^{2/3}\hspace{1mm} and \hspace{2mm}b(t) = 1
\end{equation*}
where c is a constant.
\vspace{1mm}
\noindent
We have found here, from our generalized metric (9), the well known behaviour of the scale factor a(t).\\

\vspace{1mm}
\noindent
In addition for\begin{equation*}
a(t) = ct^{2/3}\hspace{1mm} and \hspace{2mm}b(t) = 1
\end{equation*} 
we have  
\begin{equation}
{G}_{00}=\frac{4}{3t^2}\frac{(1+\frac{Gm}{2r})^2}{(1-\frac{Gm}{2r})^2}\hspace{1mm},
\end{equation}
\begin{equation}
{G}_{01}=\frac{4Gm}{3r^2(1-\frac{Gm}{2r})^2(1+\frac{Gm}{2r})^2ct^{5/3}}.
\end{equation}
We note here that the behaviour of ${G}_\text{01}$, is as $\frac{1}{t^{5/3}}$; since ${G}_\text{01}$ is a momentum component,its positivity shows that matter falls outward. This may be a fact that supports the proposition that the universe was created from a black hole.

\vspace{1mm}
From this we note that the energy due to dust calculated for any finite volume including the black hole is infinite.
In his latest talk \cite{M}, S.W.Hawking states that "black holes are not black", this can be a manifestation of what we have stated previously.

\vspace{1mm}
 Another fact that we need to state is that, as we propose it, our event horizon is not static because of b(t). Our result for dust solution is $b(t)=1$ , which implies a static event horizon. This has been pointed also by A.Davidson and S.Rubin \cite{L}: an evolving universe can host locally a static event horizon.

\subsection{The case for a diagonal Einstein Tensor}

\noindent
For ${G}_\text{01}$=0, from equation (11) we obtain simply
\begin{equation}
b(t)=\frac{1}{a(t)}\hspace{1mm}.
\end{equation}
Replacing the value of b(t) into (10) and (12), we get the following equations
\begin{equation}
{G}_{00}=3\hspace{1mm}\frac{\dot{a}^2}{a^2}\hspace{1mm},
\end{equation}
\begin{multline}
{G}_{11}={G}_{22}={G}_{33}=\\-\frac{1}{(1-\frac{Gm}{2ra(t)})} \bigg(2\frac{\ddot{a}}{a} + \frac{\dot{a}^2}{a^2} + \Big(\frac{Gm}{2ra(t)}\big(2\frac{\ddot{a}}{a} -5 \frac{\dot{a}^2}{a^2}\big)\Big)\bigg).
\end{multline}
\noindent
We notice that this is the solution produced again by our generalized metric (9). They are identical to equations (7) and (8) of the McVittie metric.\\
\vspace{1mm}

\noindent
Let us introduce the Hubble parameter:
\begin{equation}
H=\frac{\dot{a}}{a}\hspace{1mm}.
\end{equation} 
Consequently the equations (16) and (17) become
\begin{multline}
{G}_{00}=3H^2\hspace{1mm},\hspace{1mm}
{G}_{11}={G}_{22}={G}_{33}=\\-\frac{1}{(1-\frac{Gm}{2ra(t)})}\big(2\dot{H}+3H^2+\frac{Gm}{2ra(t)}(2\dot{H}-3H^2)\big).
\end{multline}
For \begin{equation}
H=\frac{\dot{a}}{a}=constant\hspace{1mm},
\end{equation} we have:
${G}_\text{00}=3H^2$ and\\

\vspace{1mm}\noindent ${G}_\text{11}={G}_\text{22}={G}_\text{33}=-3H^2$\hspace{2mm}which is independent of r.\\

\noindent
Equivalently, it can be shown that if ${G}_\text{11},{G}_\text{22},{G}_\text{33}$ are independent of r, H is constant.\\

\noindent
We find that for $b(t)=\frac{1}{a(t)}\hspace{1mm}and \hspace{1mm}H=constant$\\
\noindent
${G}_\text{00}=-{G}_\text{11}=-{G}_\text{22}=-{G}_\text{33}=3H^2$\hspace{2mm}which is constant.
This is the vacuum (dark) energy for dark energy dominated universe.

\vspace{1mm}
There are other solutions notably presented in \cite{L}, the static case limit of the perfect fluid flow and time dependent solutions assuming that the spacetime hosts a local event Killing horizon. For the latter, a perturbative analysis is conducted for the metric components, order by order, with a dimensionless radial coordinate serving as an expansion parameter, around the local Killing horizon. 

\section{Conclusion}
We have investigated the effects of conformal factors a(t) and b(t) in the background of a homogeneous and isotropic expanding universe. Consequently, field equations were presented for various cases to illustrate our point of view. From a more generalized metric that we have proposed in the beginning of this paper, we have shown that, depending on given conditions, we can obtain the characteristics of a(t) and b(t) for the matter dominated universe ($p=0$, matter dominated dust approximation) and the vacuum (dark) energy dominated universe ($p=-\rho$) with the McVittie metric.\\

Concerning the radius of the black hole, we need to make some remarks.
In our terms, the radius is $r=2Gmb(t)$. 
For the expanding universe, with the dust approximation, $b(t)=1$ we note that the radius is $a(t)r = 2Gma(t)$. This means that the black hole radius increases with the expansion as $t^{2/3}$.
In the case that $b(t)=\frac{1}{a(t)}$, the McVittie case, we have $a(t)r = 2Gm = constant$; the black hole radius is constant with the expansion.\\

In standard cosmology, for a universe with matter, $p=0$, and the energy density which is proportional to the square of the Hubble parameter varies as $a^{-3}$ and for purely dark energy with $p=-\rho$ the Hubble parameter is constant.\\

\noindent$\Lambda$CDM fit to today's expanding universe gives:
\begin{equation*}
\big(\frac{H}{H_\text{0}})^2 = 0.75 + 0.25\big(\frac{a_\text{0}}{a})^3,
\end{equation*}
where $H_\text{0}$ and $a_\text{0}$ are respectively the current values of the Hubble parameter and the scale factor.

\vspace{1mm}
Hence today's universe is 75\% dark energy dominated for which the Schwarzschild radius is constant. But due to the 25\% matter we expect the Schwarzschild radius to increase but not at a slower rate than $t^{2/3}$.\\
\newpage
\noindent 
We have given explicit solutions for pure dark energy for which for $a(t)=ce^{\sqrt{{\Lambda}/3}\hspace{1mm}t}$ and for pure matter for which $a(t)=ct^{2/3}$. However following the analysis given in \cite{L} we expect that solutions for other equations of state require a(t) to be replaced by R(r,t). 

\vspace{2mm}

\section{Acknowledgements}
Y. \c{S}eniko\u{g}lu and M. Ar{\i}k thank Prof. A.Davidson for his insights and helpful conversations that remarkably improved the presentation of the paper. 
\bibliography{mybib}{}
\bibliographystyle{unsrt}
\vspace*{-0.5cm}
\end{document}